\documentstyle[12pt]{article}
\begin{document}

\baselineskip 24pt

\newcommand{\sheptitle}
{A New Standard Parametrisation of the
Cabibbo-Kobayashi-Maskawa Matrix}

\newcommand{\shepauthor}
{S. F. King }

\newcommand{\shepaddress}
{Physics Department, University of Southampton
\\Southampton, SO17 1BJ}

\newcommand{\shepabstract}
{We propose a new parametrisation of the
Cabibbo-Kobayashi-Maskawa Matrix
in which the approximations of the standard parametrisation,
$|V_{cb}|\approx s_{23}$,
$|V_{us}|\approx s_{12}$,
are promoted to exact results.}

\begin{titlepage}
\begin{flushright}
SHEP 95-15\\
hep-ph/9505362
\end{flushright}
\vspace{.4in}
\begin{center}
{\large{\bf \sheptitle}}
\bigskip \\ \shepauthor \\ \mbox{}
\\ {\it \shepaddress} \\ \vspace{.5in}
{\bf Abstract} \bigskip \end{center} \setcounter{page}{0}
\shepabstract
\end{titlepage}

There are myriad possible
parametrisations of the
Cabibbo-Kobayashi-Maskawa (CKM) Matrix $V$ \cite{CKM}.
One particular parametrisation \cite{standard}
has recently been advocated by the
Particle Data Group (PDG) \cite{PDG}
as a ``standard parametrisation'' :

\begin{equation}
V=
\left(\begin{array}{ccc}
c_{12}c_{13} & s_{12}c_{13} & s_{13}e^{-i\delta} \\
-s_{12}c_{23}-c_{12}s_{23}s_{13}e^{i\delta} &
c_{12}c_{23}-s_{12}s_{23}s_{13}e^{i\delta} & s_{23}c_{13} \\
s_{12}s_{23}-c_{12}c_{23}s_{13}e^{i\delta} &
-c_{12}s_{23}-s_{12}c_{23}s_{13}e^{i\delta} & c_{23}c_{13}
\end{array} \right)
\label{standard1}
\end{equation}
where $s_{ij}\equiv \sin \theta_{ij}$,
$c_{ij}\equiv \cos \theta_{ij}$.

The standard parametrisation in Eq.\ref{standard1}
has certain virtues
from the point of view of $b$ quark physics,
since the physical quantities
$|V_{ub}|$,$|V_{cb}|$,$|V_{us}|$, may be
identified with simple combinations of angles,
\begin{eqnarray}
|V_{ub}| & \equiv & s_{13} \nonumber \\
|V_{cb}| & \equiv & s_{23}c_{13} \nonumber \\
|V_{us}| & \equiv & s_{12}c_{13}
\label{standard2}
\end{eqnarray}
and also, as in the Wolfenstein
parametrisation \cite{Wolfenstein},
the phase $\delta$ is always associated with the small
parameter
\begin{equation}
|V_{ub}|= s_{13}\approx 0.002-0.005.
\label{vub}
\end{equation}
Indeed the experimental smallness of this quantity \cite{PDG}
implies that
\begin{equation}
c_{13}=(1-|V_{ub}|^2)^{1/2} \approx 1
\label{approx1}
\end{equation}
to an accuracy of $\frac{1}{2}|V_{ub}|^2\approx 10^{-5}$,
or $10^{-3}\%$.
Making the approximation in Eq.\ref{approx1},
two of the results in Eq.\ref{standard2} become even simpler
\begin{eqnarray}
|V_{cb}| & \approx & s_{23} \nonumber \\
|V_{us}| & \approx & s_{12} .
\label{approx2}
\end{eqnarray}
The accuracy with which $|V_{us}|$, for example, is measured
is about 1\%, which is much larger than error incurred by
making the approximation
in Eq.\ref{approx1}. Thus the results in Eq.\ref{approx2}
are excellent approximations. Using the approximation
in Eq.\ref{approx1},
the full CKM matrix in Eq.\ref{standard1}
may be approximated by
\begin{equation}
V\approx
\left(\begin{array}{ccc}
c_{12} & s_{12}  & s_{13}e^{-i\delta} \\
-s_{12}c_{23}-c_{12}s_{23}s_{13}e^{i\delta} &
c_{12}c_{23}-s_{12}s_{23}s_{13}e^{i\delta} & s_{23} \\
s_{12}s_{23}-c_{12}c_{23}s_{13}e^{i\delta} &
-c_{12}s_{23}-s_{12}c_{23}s_{13}e^{i\delta} & c_{23}
\end{array} \right).
\label{approx3}
\end{equation}
We emphasise that Eq.\ref{approx3} is adequate for all
practical purposes, including CP violation.
For example the CP-violating quantity $J_{CP}$ \cite{J}
calculated using Eq.\ref{standard1}
has the excellent approximation
when calculated using Eq.\ref{approx3},
\begin{equation}
J_{CP}=c_{13}^2c_{12}c_{23}\sin\delta s_{12}s_{23}s_{13}
\approx c_{12}c_{23}\sin\delta s_{12}s_{23}s_{13}.
\end{equation}

In this Brief Report we introduce a
parametrisation which is similar
to the standard parametrisation in Eq.\ref{standard1}, but which
promotes the desirable approximate results in Eq.\ref{approx2}
to exact relations. Our motivation is simply that
we find it more physically appealing
to regard the relations in Eq.\ref{approx2}
as exact. To this end we replace $\theta_{23}$, $\theta_{12}$ by
two new angles $\theta'_{23}$,
$\theta'_{12}$, and retain the original definition of $\theta_{13}$.
The new angles are defined by the relations:
\begin{eqnarray}
|V_{ub}| & \equiv & s_{13} \nonumber \\
|V_{cb}| & \equiv & s_{23}' \nonumber \\
|V_{us}| & \equiv & s_{12}' .
\label{new1}
\end{eqnarray}
Comparing Eq.\ref{new1} to Eq.\ref{standard2} we see that
the primed angles are given by a
simple change of variables,
\begin{eqnarray}
s_{23}'& \equiv & s_{23}c_{13} \nonumber \\
s_{12}'& \equiv & s_{12}c_{13}
\label{variables}
\end{eqnarray}
As in the standard parametrisation we eliminate the phases in the
elements $V_{ud}$,$V_{us}$,
$V_{cb}$,$V_{tb}$,
and introduce the standard phase
into $V_{ub}$. Eq.\ref{new1} then implies,
\begin{equation}
V=
\left(\begin{array}{ccc}
\sqrt{c_{13}^2-s_{12}'^2} & s_{12}' & s_{13}e^{-i\delta} \\
V_{cd} & V_{cs} & s_{23}' \\
V_{td} & V_{ts} & \sqrt{c_{13}^2-s_{23}'^2}
\end{array} \right)
\label{new2}
\end{equation}
The complex CKM matrix elements
$V_{cd}$, $V_{cs}$, $V_{td}$, $V_{ts}$,
may be determined by the unitarity
conditions of rows and columns,
or alternatively using the change of
variables in Eq.\ref{variables},
\begin{eqnarray}
V_{cd} & = & -\frac{s_{12}'\sqrt{c_{13}^2-s_{23}'^2}}
                            {c_{13}^2}
 -\frac{\sqrt{c_{13}^2-s_{12}'^2}s_{23}'s_{13}e^{i\delta}}
                            {c_{13}^2} \nonumber \\
V_{cs} & = & \frac{\sqrt{c_{13}^2-s_{12}'^2}
           \sqrt{c_{13}^2-s_{23}'^2}}
                               {c_{13}^2}
                  -\frac{s_{12}'s_{23}'s_{13}e^{i\delta}}
                               {c_{13}^2} \nonumber \\
V_{td} & = & \frac{s_{12}'s_{23}'}
                    {c_{13}^2}
 -\frac{\sqrt{c_{13}^2-s_{12}'^2}
\sqrt{c_{13}^2-s_{23}'^2}s_{13}e^{i\delta}}
                               {c_{13}^2}   \nonumber \\
V_{ts} & = & -\frac{\sqrt{c_{13}^2-s_{12}'^2}s_{23}'}
                            {c_{13}^2}
 -\frac{s_{12}'\sqrt{c_{13}^2-s_{23}'^2}s_{13}e^{i\delta}}
                            {c_{13}^2}
\label{new3}
\end{eqnarray}
If we make the approximation in Eq.\ref{approx1},
then the new parametrisation in Eqs.\ref{new2},\ref{new3}
reduces to the approximate form in Eq.\ref{approx3}.
This is to be expected since under the approximation
in Eq.\ref{approx1} the primed and unprimed variables
 in Eq.\ref{variables} are equivalent. Thus
there is no practical difference between the standard
parametrisation in Eq.\ref{standard1}, and the new
parametrisation in Eqs.\ref{new2},\ref{new3},
since both reduce to Eq.\ref{approx3} under
the excellent approximation in Eq.\ref{approx1}.
However the new parametrisation
has the aesthetic virtue that
the approximations of the standard parametrisation,
$|V_{cb}|\approx s_{23}$,
$|V_{us}|\approx s_{12}$,
are promoted to exact results.

\begin{center}

{\bf Acknowledgements}

\end{center}

I would like to thank Ken Barnes and
Jon Flynn for conversations.

\end{document}